# BSSSN: Bit String Swapping Sorting Network for Reversible Logic Synthesis

Md. Saiful Islam
Dept. of Computer Science & Engineering
State University of Bangladesh
Email: sohel_csdu@yahoo.com

**Abstract**

In this paper, we have introduced the notion of UselessGate and ReverseOperation. We have also given an algorithm to implement a sorting network for reversible logic synthesis based on swapping bit strings. The network is constructed in terms of n*n Toffoli Gates read from left to right and it has shown that there will be no more gates than the number of swappings the algorithm requires. The gate complexity of the network is $O(n^2)$. The number of gates in the network can be further reduced by template reduction technique and removing UselessGate from the network.

**Major Area:** VLSI, Logic Synthesis & Reversible Logic.

## 1. Introduction

Power dissipation is a very important factor in VLSI design. The main benefit of reversible logic is theoretically zero power dissipation. A reversible circuit maps each input vector into a unique output vector. Landaur's principle [1] proved that logic computations that are not reversible necessarily dissipate heat irrespective of their implementation technologies. According to [2] zero energy dissipation would be possible only if the network consists of reversible gates. Thus reversibility will become an essential property in future circuit design.

Synthesis of reversible logic circuits differs significantly from the synthesis of combinational logic circuits. Because in a reversible circuit the number of inputs must be equal to the number of outputs, every output can be used only once (i.e., no fan-out is permitted), and must be acyclic.

Although there exist many reversible gates in the literature good synthesis methods have not yet emerged. Miller *et al* have proposed transformation based algorithm with template reduction method in [5][6] and spectral techniques in [7] to find near optimal circuits. Mischenko and Perkowski in [8] have suggested a regular structure of reversible wave cascades and show that such a structure requires no more than the product terms in an ESOP realization of the function. A regular symmetric structure has been proposed by Perkowski *et al*. [9] to realize symmetric functions. Sorting Network for reversible logic synthesis has been described in [10]. The idea here is used based on swapping bit strings. In fact one would expect that a better method could be found.

In this paper, we have presented a convergent synthesis algorithm BSSSN and a variant of its Var_BSSSN based on swapping bit strings. The algorithm constructs a network of n*n Toffoli Gates read from left to right both for input and output translation. We have also given proofs why our algorithm will converge and introduced the notion of UselessGate and ReverseOperation that were not discussed previously. The paper is organized as follows: Section 2 provides some definition and basics necessary to understand our algorithm, Section 3 describes our algorithm, Section 4 explains how the network can be refined to reduce its circuit width, Section 5 will conclude the paper and references are cited in Section 6.

## 2. Background

An *n*-input *n*-output totally specified Boolean function $f(X)$, $X = \{x_1, x_2, \ldots, x_n\}$ is reversible iff it maps each input assignment to a unique output assignment.

A reversible function can be written as a standard truth table as in Table 1 and can also be viewed as a bijective mapping of the set of integers $0, 1, \ldots, 2^n-1$. Hence a reversible function can be defined as an ordered set of integers corresponding to the right side of the table, e.g. {1,0,3,2,5,7,4,6} for the function in Table 1. We can thus interpret the function over the integers as $f(0) = 1$, $f(1) = 0$, $f(2) = 3$, etc.

An *n*-input *n*-output gate is reversible if it realizes a reversible function. Many reversible gates have been proposed in the literature. One of the first gates was the CNOT gate [3], which capable of producing the "exclusive or" of two input bits as the second output and the first output is equal to the first input.

| c | b | a | c° | b° | a° |
|---|---|---|----|----|----|
| 0 | 0 | 0 | 0  | 0  | 1  |
| 0 | 0 | 1 | 0  | 0  | 0  |
| 0 | 1 | 0 | 0  | 1  | 1  |
| 0 | 1 | 1 | 0  | 1  | 0  |
| 1 | 0 | 0 | 1  | 0  | 1  |
| 1 | 0 | 1 | 1  | 1  | 1  |
| 1 | 1 | 0 | 1  | 0  | 0  |
| 1 | 1 | 1 | 1  | 1  | 0  |

**Table 1. 3*3 Reversible logic function**

A generalization of CNOT is a 3-input 3-output Toffoli gate [4]. The Toffoli gate negates the third bit iff the first two bits are 1. Figure 1 shows both gates as they are commonly drawn.

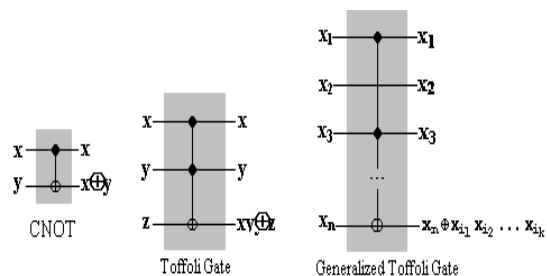

**Fig 1.  CNOT & Toffoli Gate**     **Fig 2. Generalized Toffoli Gate**

A generalized $n*n$ Toffoli gate changes one bit, called the target, if some of the *k* bits are 1 which is shown in Figure 2. The changing bit, also called target, may also be in any position. The gate will be defined as follows $T(x_{i1}, x_{i2}, \ldots, x_{ik} : x_n)$ where $x_n$ is the target and $x_{i1}, x_{i2}, \ldots, x_{ik}$ are the control bits.

Garbage is the number of outputs added to make an *n*-input *k*-output function ($(n, k)$ function) reversible.

Given two bit strings, *P* and *Q*, the Hamming distance between them, denoted by $\delta(P, Q)$ is the number of positions for which *P* and *Q* differ.

*Example 1.* Consider the bit strings (1,0,1) and (0,1,1). The Hamming distance between these two bit strings is 2 since the number of positions for which these two bit strings differ is 2.

Given the function $f(X)$, the complexity $C(f)$ is defined as the sum of the individual Hamming distances over the $2^n$ input-output patterns. For example, the value of $C(f)$ for the function in Table 1 is 8.

***Lemma 1.*** The upper and lower bound on the Hamming distance $\delta$, between any two bit strings P and Q, in any reversible specification is $n$ and 1, where $n$ is the number of input lines. That is,

$$1 \leq \delta(P, Q) \leq n.$$

*Proof:* Let P be $(a_1, a_2, \ldots, a_n)$ and Q be $(b_1, b_2, \ldots, b_n)$. Since in a reversible specification no two bit strings are identical, they must differ in at least one position. Let $m$ be the index at which $a_m \neq b_m$. That is, $a_m = b_m{'}$, and $b_m$ is either 0 or 1. Bit strings $P$ and $Q$ may differ at most every position, i.e., $a_i \neq b_i$, where $1 \leq i \leq n$. Thus we can conclude that $1 \leq \delta(P, Q) \leq n$.

***Lemma 2.*** Two bit strings $P$ and $Q$ can be swapped without affecting others iff the Hamming distance between them is 1.

*Proof:* In any reversible specification bit string P and Q will be unique. Let P be $(p_1, p_2, \ldots, p_n)$ and Q be $(q_1, q_2, \ldots, q_n)$. Also let m be the index for which $p_m \neq q_m$. That is, $p_i = q_i$, where $1 \leq i \leq n$ and $i \neq m$. Since no bit string except $P$ & $Q$ will contain $(p_1, p_2, \ldots, p_{m-1}, x, p_{m+1}, \ldots p_n)$ where $x$ may be either '0' or '1'. If we use $(p_1, p_2, \ldots, p_{m-1}, p_{m+1}, \ldots p_n)$ as the control bits to drive the Toffoli gate, only P & Q will be affected.

***Example 2.*** In Figure 3, bit strings (1,1,1) and (1,1,0) have been swapped using $T(b, c: a)$. Since the Hamming distance between them is one, this swapping can be carried out without affecting others.

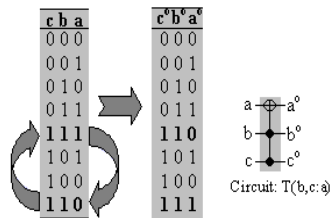

**Fig 3. Swapping bit strings**

**UselessGate**: A pair of gates in a network is considered to be useless if it has no effect in the circuit, i.e., when it is identified and removed from the network it will have no effect in the final output produced by the circuit.

***Example 3.*** $T(c, b : a)$ is an example of a useless gate in the gate sequence $T(c, b : a)\ T(c : a)\ T(c, b : a)$ and the gate sequence can be replaced by $T(c : a)$ without any change or modification in the circuit. The final output produced by the new circuit will be the same as the previous one.

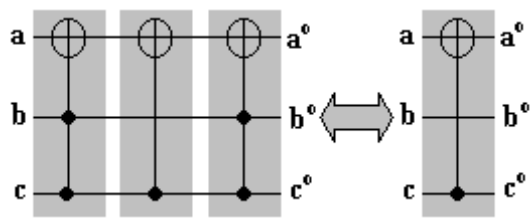

**Fig 4: Example of UselessGate**

To identify and then to remove a UselessGate, we take into account the following property which follows directly from the definition of $n*n$ Toffoli gates.

***PROPERTY 2.1:*** A gate $T(x_1, x_2, \ldots, x_{k-1} : x_k)$ can be removed from the sequence $T(x_1, x_2, \ldots, x_{k-1} : x_k)\ T(a_1, a_2, \ldots, a_{l-1} : a_l)\ T(b_1, b_2, \ldots, b_{m-1} : b_m) \ldots T(c_1, c_2, \ldots, c_{n-1} : c_n)\ T(x_1, x_2, \ldots, x_{k-1} : x_k)$ iff $x_k \notin \{a_1, a_2, \ldots, a_{l-1}, b_1, b_2, \ldots, b_{m-1}, \ldots, c_1, c_2, \ldots, c_{n-1}\}$ and $a_l, b_m, \ldots, c_n \notin \{x_1, x_2, \ldots, x_{k-1}\}$.

**ReverseOperation**: A reverse operation places a bit string Z to its previous place if it is transferred to a different place by another operation(s).

Consider a sequence of operations $\ldots S_l(Y, Z) \ldots S_k(X, Z) \ldots S_m(Z, Y) \ldots$ in which $S_m$ is the reverse operation of $S_l$, where operation $S$ swaps its arguments.

***Example 4***: In the example below let we want to swap (1,1,0) and (1,0,1). But we cannot do it directly since the Hamming distance between them is greater than one. This can be carried out by a sequence of swap operations: first (1,0,1) and (1,1,1), then (1,1,0) and (1,1,1). Finally (1,1,1) and (1,0,1).

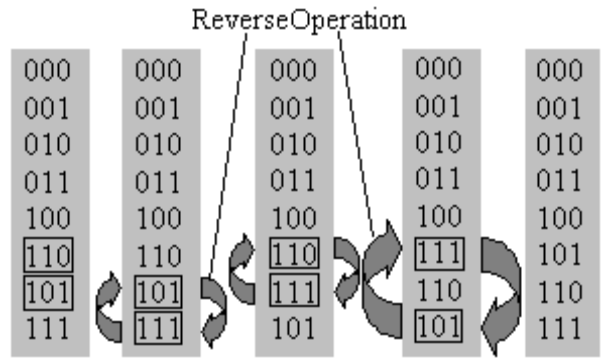

**Fig 5: ReverseOperation**

## 3. Sorting Network

Section 2 has described that a reversible function can be defined as an ordered set of integers corresponding to the right side of the table, e.g. {1,0,3,2,5,7,4,6} for the function in Table 1. Therefore, if we can build a network of reversible gates that might sort this set, it will eventually realize the function. For example, the ordered set of integers for the function in table 1 will become {0,1,2,3,4,5,6,7} and the index of each element will be equal to itself. That is, if we define the set as $\{p_i\}$, then $2^n-1 \geq i \geq 0$ and $p_i = i$ where n is the number of input lines. Section 3.1, 3.2, 3.3 and 3.4 will present such an algorithm named BSSSN (Bit String Swapping Sorting Network) that will construct a network as a sequence of Toffoli gates that might sort the elements of the set. The idea here is based on swapping bit strings whose hamming distance is exactly one as described in section 2.

### 3.1. Output translation

**BSSSN:** Constructing the sorting network and
its corresponding circuit

While (there is a bit string $(a_1,a_2,\ldots,a_n)$ in the set that is not in its intended place, i.e., int_value$(a_1,a_2,\ldots,a_n) \neq$ index)

    Step 1. Let $(a_1,a_2,\ldots,a_n)$ is not in its intended place then by induction its place is occupied by another string, say $(b_1,b_2,\ldots,b_n)$
    Step2. Compute dist = $\delta\ ((a_1,a_2,\ldots,a_n), (b_1,b_2,\ldots,b_n))$
        if dist=1, swap $(a_1,a_2,\ldots,a_n)$ and $(b_1,b_2,\ldots,b_n)$ using gate $T(a_1,a_2,\ldots, a_{k-1},a_{k+1}, \ldots,a_n : a_k)$
        where $a_k \neq b_k$
        else{
            find all the bit strings $\{(c_1,c_2,\ldots,c_n)\}$ such that $\delta\ ((b_1,b_2,\ldots,b_n), (c_1,c_2,\ldots,c_n))=1$
            swap $(b_1,b_2,\ldots,b_n)$ with one of the$(c_1,c_2,\ldots,c_n)$ such that $\delta\ ((a_1,a_2,\ldots,a_n), (c_1,c_2,\ldots,c_n))$
            will be minimum and in case of multiple $(c_1,c_2,\ldots,c_n)$ low int_value bit string is
            chosen to break the tie and that is not in its intended place

            if $(c_1,c_2,\ldots,c_n)$ is in its intended place then do the ReverseOperation
        }
    Step3. Now $(c_1,c_2,\ldots,c_n)$ will become the new $(b_1,b_2,\ldots,b_n)$ and goto step 2.

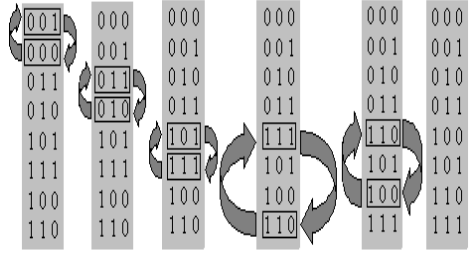
**Fig 6. Constructing network using BSSSN**

Figures 6 and 7 illustrate the application of BSSSN. The sequence of gates that form the network is: T(b′,c′:a) T(b,c′:a)T(a,c:b)T(b,c:a) T(a′,c:b).

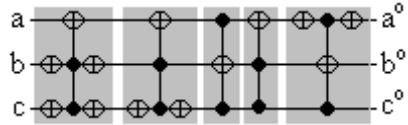
**Fig 7. Circuit for the network constructed by BSSSN**

Algorithm BSSSN is straightforward. It is greedy in the sense it hopes that a bit string can be swapped by the one that is in its intended place. Because of Lemma 1 and 2, it is always possible to find two bit strings for Steps 2 that can be swapped. Therefore, it will always terminate successfully with a circuit for a given specification. The best case occurs when a bit string can always be swapped by the bit string placed in its intended position. A variant of BSSSN takes the bit string in the set whose integer representation is low and brings it to its intended place. Figures 8 and 9 illustrate the application of the variant of BSSSN and the sequence of gates that form the network is: T(b′,c′:a)T(b,c′:a)T(b′,c:a)T(a,c:b) T(b,c:a).

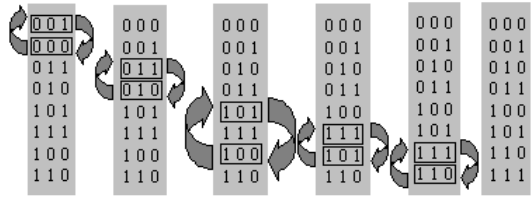
**Fig 8. Constructing network using variant of BSSSN**

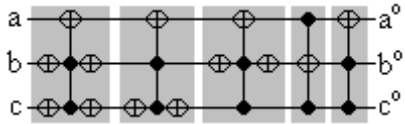
**Fig 9. Circuit for the network constructed by variant of BSSSN**

### 3.2 Input Translation

For input translation, we have to find an inverse of the specification. For example, the reverse specification of the function in Table 1 is {1,0,3,2,6,4,7,5}. Then, we can apply BSSSN to realize the function. By applying BSSSN to realize the reverse specification we get the circuit: T(b′,c′:a)T(b,c′:a)T(b,c :a)T(a,c:b)T(b′, c:a) and its variant, we get: T(b′,c′:a)T(b, c′:a)T(a′,c:b) T(b,c:a)T(a,c:b)T(b,c:a).

### 3.3 Random Selection and Control input Reduction

To sort the elements in a specification, algorithms BSSSN and its variant take one element at a time and bring it to its intended place. Selection of a bit string randomly and then returning to its intended place can reduce the total number of swapping. This will minimize the total number of gates in the network also.

Algorithm BSSSN also assigns the maximum number of control lines to each Toffoli gate. For larger problems with up to 8 or 9 inputs this may not be a practical one. Selective use of control inputs can be used to swap elements. This can be carried out safely as long as it will not affect bit strings that are already in its intended place. We should choose a subset of the control inputs that will minimize the $C(f)$ of the resulting specification. For example, we can select control inputs that will drive a Toffoli gate to bring more than one element at a time to their intended place.

| Function $f$ | | Function $f^{-1}$ | |
|---|---|---|---|
| c b a | c°b°a° | c°b°a° | c b a |
| 0 0 0 | 0 0 1 | 0 0 0 | 0 0 1 |
| 0 0 1 | 0 0 0 | 0 0 1 | 0 0 0 |
| 0 1 0 | 0 1 1 | 0 1 0 | 0 1 1 |
| 0 1 1 | 0 1 0 | 0 1 1 | 0 1 0 |
| 1 0 0 | 1 0 1 | 1 0 0 | 1 1 0 |
| 1 0 1 | 1 1 1 | 1 0 1 | 1 0 0 |
| 1 1 0 | 1 0 0 | 1 1 0 | 1 1 1 |
| 1 1 1 | 1 1 0 | 1 1 1 | 1 0 1 |

**Fig 10. Reverse Specification**

### 3.4 Reduction Rules

The circuits produced by BSSSN as described thus far frequently have gate sequences that can be reduced. For example, the sequence T(b:a) T(:b) T(:a) can be replaced by the sequence $T(:b)$ T(b:a). Here we have implemented template driven reduction method introduced in [5]. In addition to template matching in [5] we have also removed useless gates that come in pairs and have no effect in the circuit.

### 4. Experimental Results

In Section 3.1 we have shown some reversible examples and compare them with the circuits in [6][7]. Section 3.2 describes the method used to convert an irreversible specification to a reversible one and the way to synthesize them. Here we have synthesized a reversible circuit from an irreversible specification and not transformed an irreversible circuit to a reversible one.

### 3.1. Reversible examples

For each example, the specification is given as an ordered set of integers, which define the truth table specification of the reversible logic function to be realized. The circuit is given as an ordered sequence of Toffoli gates. Read from left to right they transform the left side to the right side.

**Example 3.1** Verification of realizing a **Fredkin gate.** This example is collected from [7]. The circuit given by our method produces the same result.
*Specification:* {0,1,2,3,4,6,5,7}
*Circuit(BSSSN)* **:** T(a**:**b) T(b,c**:**a)T(a**:**b)
*Circui(VAR_ BSSSN)* **:** T(b**:**a)T(a,c**:**b)T(b**:**a)

**Example 3.2** This is a second example of the interchange of two positions in the specification. The circuit given by our method is identical to the solution provided by [8].
*Specification:* {0,1,2,4,3,5,6,7}
*Circuit(BSSSN)* **:** T(a,b**:**c) T(a,c**:**b) T(b′,c**:**a) T(a,c**:**b) T(a,b**:**c).
*Circui(VAR_ BSSSN)***:**  T(a′,b′**:**c)T(b′,c′**:**a)T(a,c′**:**b) T(b′, c′**:**a) T(a′,b′**:**c).

**Example 3.3** This example is taken from [6]. The circuit given by our method is identical to the solution provided by the Bidirectional Algorithm in [6].
*Specification:* {7,0,1,2,3,4,5,6}
*Circui(VAR_ BSSSN)***:** T(a,b**:**c)T(a**:**b)T(**:**a).

## 3. 2. Non-reversible examples

According to [9] an irreversible function can be realized using reversible gates with the addition of some number of constant inputs and 'garbage' outputs. The minimum number of garbage outputs required to transforming an irreversible function to a reversible one is $\lceil \log_2^m \rceil$, where m is the maximum number of times a single output pattern appears in the specification.

A single-output or a multi-output function $f$ involving input variables $x_1, x_2, \ldots, x_n$ can be transformed to a reversible specification in the following way.

  a. Compute $m$, where $m$ is the maximum output pattern multiplicity of the irreversible specification. Let $p = \lceil \log_2^m \rceil$ and $k$ is the number of outputs of $f$. The value of $k$ will be one for single-output function.
  b. If $(p + k) > n$
      i. Add $(p + k - n)$ new input variable $x_{n+1}, x_{n+2}, \ldots, x_{p+k}$ and set each of them to zero on input in the circuit.
      ii. Add n outputs each equal to one of the original inputs $x_1, x_2, \ldots, x_n$.
      iii. $k$ outputs will be realized on $x_{p+1}, x_{p+2}, \ldots, x_{p+k}$.
      iv. for single-output function replace $f$ by $f \oplus x_{n+1}$.
  Else
      i. Add $n-k$ outputs each equal to one of the original inputs $x_1, x_2, \ldots, x_{n-k}$.
      ii. The $k$ outputs will be realized on inputs $x_{n-k+1}, x_{n-k+2}, \ldots, x_n$.

It is easily verified that the specification constructed in this way maps an input pattern to a unique output pattern and is therefore reversible. The approach used by Miller *et al.* [7] adds unnecessary constant inputs and thus produces extra garbage outputs that are not actually needed to make the specification reversible.

**Example 3.4** This procedure for transforming a single-output function is illustrated for the example of the 2-input EX-OR function in Table 2. The resulting circuit is the single gate $T(a:b)$ which realizes the EX-OR of $a$ and $b$ on $b$. But the same function is realized in [7] with one constant on inputs and thus produces one extra garbage output that is unnecessary.

| b a | f |   | b a |   | b° a° |   | c b a |   | c° b° a° |
|-----|---|---|-----|---|-------|---|-------|---|----------|
| 0 0 | 0 |   | 0 0 |   | 0 0   |   | 0 0 0 |   | 0 0 0    |
| 0 1 | 1 |   | 0 1 |   | 1 1   |   | 0 0 1 |   | 1 0 1    |
| 1 0 | 1 |   | 1 0 |   | 1 0   |   | 0 1 0 |   | 1 1 0    |
| 1 1 | 0 |   | 1 1 |   | 0 1   |   | 0 1 1 |   | 0 1 1    |
|     |   |   |     |   |       |   | 1 0 0 |   | 1 0 0    |
|     |   |   |     |   |       |   | 1 0 1 |   | 0 0 1    |
|     |   |   |     |   |       |   | 1 1 0 |   | 0 1 0    |
|     |   |   |     |   |       |   | 1 1 1 |   | 1 1 1    |
| (a) |   |   | (b) |   |       |   | (c)   |   |          |

**Table 2.** (a) 2-input EX-OR (b) reversible specification derived from 2-input EX-OR using method described above (c) using method in [7]

**Example 3.5** This example illustrates the realization of 2-input AND function. The specification is {0,1,2,7,4,5, 6,3}. The solution produced by our algorithm is same to the solution provided by that of the [7], that is, T(a,b:c) which realizes the AND of $a$ and $b$ when $c$ is 0 on input.

**Example 3.6 Full Adder Minimization:** The resulting circuit produced by our algorithm is: T(a,b′,d′:c)T(a′,b,d′:c) T(a,b,c′:d) T(a,b′,d:c) T(a,b′,c:d)T(a′,b,d:c) T(a′,b,c:d)T(a,b,c:d).

which can be simplified by template matching to T(:d) T(a,b,d:c) T(a,d:c) T(a,b,d:c) T(b,d:c) T(:d) T(a,b,c:d)T(a, b:d)T(a,b,d:c)T(a,d:c)T(a,b ,c:d)T(a,c:d)T(a,b,d:c)T(b,d:c) (a,b,c:d)T(b,c:d)T(a,b,c:d).

which can be again simplified by removing useless gates to T(:d)T(a,d:c) T(b,d:c)T(:d)T(a,b,c:d)T(a,b:d) T(a,b,d:c) T(a,d:c)T(a,b,c:d)T(a,c:d)T(a,b,d:c)T(b,d:c) T(b,c:d).

After final simplification we get the following circuit:

T(a,b**:**d)T(a**:**b)T(b,c**:**d)T(b**:**c).

This circuit is identical to the circuit in [6][7].

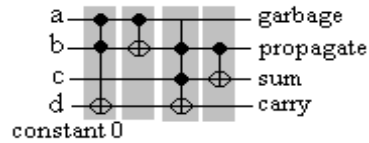
**Fig 11. Full Adder**

Though the initial circuit produced by our algorithm seems to be larger one, it can be simplified easily by simple template matching and identifying useless gates. Thus the circuit will be optimal. The main advantage of our algorithm is that it does not require exhaustive analysis like spectral used in [7].

## 5. Conclusions

In this paper we have given a convergent synthesis algorithm for reversible logic circuits. The specification must be totally specified. We have also given a method that will transform an irreversible specification to a reversible one. The algorithm always terminates with a network of Toffoli gates that can translate both input and output side to their corresponding output and input side. Since the synthesis of reversible circuits can be done in either side, this is valid.